\documentclass[aps,pre,preprint,groupedaddress,superscriptaddress,showpacs,a4paper,floatfix]{revtex4-1}

\usepackage{amsmath,amssymb,graphicx,graphics,bm}
\usepackage{hyperref}
\usepackage{graphics,graphicx}
\usepackage{amsmath,amssymb}

\begin{document}

\title{Dynamic fluctuation-dissipation theory for Generalized Langevin Equations: constructive constraints, stability and realizability}

\author{Massimiliano Giona, Giuseppe Procopio and  Chiara Pezzotti}
\email[corresponding author:]{massimiliano.giona@uniroma1.it}
\affiliation{Dipartimento di Ingegneria Chimica, Materiali, Ambiente La Sapienza Universit\`a di Roma\\ Via Eudossiana 18, 00184 Roma, Italy}

\date{\today}

\begin{abstract}
Using the initial-value formulation, a dynamic
theory for systems evolving according to a Generalized Langevin Equation
is developed, providing more restrictive  conditions
on the existence of equilibrium behavior and its fluctuation-dissipation implications.
For systems  fulfilling the property of local realizability,
 that for all the practical purposes corresponds to
the postulate of the existence of a Markovian embedding, 
physical constraints, expressed in the form of  dissipative stability
and stochastic realizability are derived.
If these two properties are met, Kubo theory is constructively recovered, while
if one of these conditions is violated a thermodynamic equilibrium behavior does not
exist (and this  occurs also for `` well-behaved dissipative
systems''  according to the classical Kubo theory), with significant implications in the linear response theory.
\end{abstract}

\maketitle

\noindent

{\bf Introduction - } The fluctuation-dissipation (FD) 
theorems developed by Kubo for Generalized Langevin Equations (GLE)  \cite{kubo1,kubo2} 
constitute a cornerstone in statistical
physics admitting a huge variety of applications in all the branches
of physics,
microhydrodynamics and transport theory, electrochemistry and dieletric response,
abstract linear response theory, etc. \cite{hydro1,gen1,gen2}, and cellular biology \cite{cell1,cell2},
  involving thermal fluctuations.

In its original formulation \cite{kubo1}, the Kubo theory addresses the generalization
of the equation of motion of a spherical Brownian particle of mass $m$
and velocity $v(t)$ in a fluid at thermal equilibrium
(constant temperature $T$), in which the dissipative response of the fluid
is not instantaneous but is described via a memory kernel $h(t)$
\begin{equation}
m \, \frac{d v(t)}{d t} = - \int_{-\infty}^t h(t-\tau) \, v(\tau) \, d \tau
+ R(t)
\label{eq1_1}
\end{equation}
where $R(t)$ is the stochastic termal force, such that the resulting velocity
process $v(t)$ is stationary.
Another basic assumption for $R(t)$ is the Langevin condition, originally
proposed by P. Langevin in 1908 \cite{langevin}, stating that
\begin{equation}
\langle R(t) \, v(\tau) \rangle_{\rm eq} =0  \qquad  \tau \leq t
\label{eq1_2}
\end{equation}
where $\langle \cdot \rangle_{\rm eq}$ is the expected value with
respect to the probability measure of the thermal fluctuations at equilibrium.
Eq. (\ref{eq1_2}) implies that the stochastic force 
$R(t)$ at time $t$ is independent
of the previous history of particle velocity. An alternative view to eq. (\ref{eq1_2})
has been developed in \cite{felder} (see also \cite{bala} for a critical
discussion).

The Kubo FD theory is developed in  two steps: the
determination of the velocity autocorrelation function $C_{vv}(t)=
\langle v(t+\tau) \, \langle v(\tau) \rangle_{\rm eq}  = \langle v(t) \, v(0) \rangle_{\rm eq}$
(due to stationarity), and this is referred to as the fluctuation-dissipation
theorem of the first kind (FD1k, for short), and the determination
of the autocorrelation function $C_{RR}(t)= \langle R(t) \, R(0) \rangle$
for the thermal force, referred to as the fluctuation-dissipation
theorem of the second kind (FD2k, for short). 
The development of the Kubo theory starts from eq. (\ref{eq1_1})
and involves essentially Fourier analysis, specifically
the application of the Wiener-Khinchin theorem for stationary
stochastic processes, imposing: (i) the
Langevin condition eq. (\ref{eq1_2}) and (ii) the value
of the intensity of velocity fluctuations at thermal
equilibrium,
\begin{equation}
\langle v^2 \rangle_{\rm eq}= \frac{k_B \, T}{m}
\label{eq1_3}
\end{equation}
where $k_B$ is the Boltzmann constant.
We prefer to use  the wording ``properties'' instead of
''theorems'', as  FD1k and FD2k are 
two ``properties''  based upon
specific physical assumptions (valid for some systems, but that can be
equally well be violated by others), rather that propositions involving
mathematical entities.

It is possible to frame the same problem in a slightly different way,
 replacing eq. (\ref{eq1_1}) with
\begin{equation}
m \, \frac{d v(t)}{d t} = - \int_{0}^t h(t-\tau) \, v(\tau) \, d \tau
+ R(t)
\label{eq1_4}
\end{equation}
equipped with the initial condition $v(0)=v_0$.
The comparison of these two equations  shows
that the only difference  between them refers to the lower integration value,
that  is $\tau=-\infty$ in eq. (\ref{eq1_1}) that is turned to $\tau=0$
in  eq. (\ref{eq1_4}).
 Probably because of the apparent ``tininess'' of this
 difference, no great
attention has been focused on it,  and in the
literature these two formulations are used interchangeably
depending upon practical  convenience \cite{lit1,lit2,lit3,lit4}.
Henceforth we refer to eq. (\ref{eq1_1}) as the {\em abstract formulation}
of the GLE and to eq. (\ref{eq1_4}) as its {\em initial-value formulation}.

In point of fact, the mathematical structure of these two formulations
is different (although intrinsic analogies exist \cite{jap,jap1}), and more
importantly this difference  admits  relevant implications as regards the
methodological way of developing FD theory and the results following from it.
The abstract formulation is essentially focused at determining the statistical
properties of the stochastic processes $v(t)$ and $R(t)$
 at equilibrium.
 In this way, it fits
the purposes of an equilibrium theory for the linear response of physical
systems to perturbations (Kramers-Kr\"onig relations, analysis
of susceptibilities, etc.) \cite{kubo2,degroot}.
As it represents intrinsically an equilibrium formulation,  it is
unable to handle
neither  the momentum relaxation dynamics and the convergence towards 
mechanical equilibrium
nor   any experiments in which
a particle, possessing initially a velocity $v_0$,   is injected
into the fluid at some
initial time, say $t=0$.
Conversely, the initial-value formulation is a dynamic, non-equilibrium
(as regards momentum transfer, not thermal effects)
description of the process, i.e. of the hydromechanical interactions between the particle and the fluid. It corresponds to the
description of experiments involving a particle (or a system of independent particles),
injected into the fluid and relaxing its momentum dynamics towards equilibrium
conditions,
 providing a stochastic evolution equation for the
particle velocity,  the results of which can be directly checked against experiments such as those involving Brownian motion \cite{exp0,exp1,exp2,exp3}. 
For the sake of physical correctness, eq. (\ref{eq1_4})  describes the fluid-particle
interactions in a viscous fluid neglecting fluid-inertial effects
\cite{franosch,procopiovisco}, in which the  memory kernel
$h(t)$ accounts for the   viscoelastic dissipative properties of the fluid \cite{rheol1,rheol2}.

The Kubo FD theory has been developed starting from eq. (\ref{eq1_1})
and there is no direct proof of  it starting from eq. (\ref{eq1_4}). The
present Letter is aimed at  developing
the necessary technical tools for a FD theory starting from eq. (\ref{eq1_4}), and derive
the physical implications.  This change of perspective  permits to
derive new and apparently unexpected  properties of FD theory, associated with its physical realizability,
existence of equilibrium properties, and ultimately applicability of the Stokes-Einstein relation  for
simple and ``well-behaved'' dissipative memory kernels.
The concept of local realizability of the memory kernel,  dissipative stability of the GLE, and stochastic realizability 
for the  thermal fluctuations 
are introduced as necessary constraints corresponding to thermodynamic consistency conditions.
Depending on their fulfillment several  different regimes are observed and explained.
The results obtained are not only consistent with the mathematical theory of linear integrodifferential
equations \cite{math1,math2}, but also show unexpected features once compared and contrasted against the classical FD theory: 
in apparently ``simple  well-behaved dissipative'' systems (see further for a definition),
for which the classical Kubo theory predicts a regular diffusive behavior satisfying Stokes-Einstein relations,
the present theory correctly predicts and explains the violation of the Stokes-Einstein relation as a consequence of
the lack of any thermodynamic equilibrium behavior. Moreover, the present theory provides a simple interpretation
of the phenomena associated with ergodicity breaking of simple dissipative GLE in the presence of  memory kernels exponentially
decaying in time \cite{bao1,bao2,bao3,plyukhin1,plyukhin2}. For a thorough analysis of this case see \cite{ergbreak}.
The present approach  constitutes a  necessary complement and a significant refinement of  the linear response theory as regards
thermodynamic consistency of the memory kernels and of their interaction with equilibrium
fluctuations.

\noindent

{\bf FD1k  and FD2k -}  To begin with, consider FD1k.
This follows directly from eq. (\ref{eq1_4}) as a consequence of
the Langevin condition eq. (\ref{eq1_2}), multiplying it by $v(0)$ and taking
the expected value with respect to the equilibrium probability measure:

\begin{equation}
m \, \frac{d \langle v(t) \, v(0) \rangle_{\rm eq}}{d t}= m \, \frac{d C_{vv}(t)}{d t}
= -\int_0^t h(t-\tau) \, C_{vv}(\tau) \, d \tau
\label{eq2_5}
\end{equation}
equipped with the initial condition $C_{vv}(0)=\langle v^2 \rangle_{\rm eq}$.
In the Laplace domain, setting $\widehat{h}(s)= L[h(t)]=\int_0^\infty e^{-s t} \, h(t) \, dt$,
and similarly for the other functions, the solution of eq. (\ref{eq2_5}) is 
\begin{equation}
\widehat{C}_{vv}(s)= \frac{\langle v^2 \rangle}{s + \widehat{h}(s)/m}
=  \langle v^2 \rangle \, \widehat{G}(s)
\label{eq2_6}
\end{equation}
where $\widehat{G}(s)$ is the Laplace transform of the Green
function (referred to as the resolvent in \cite{math1} associated with the memory dynamics defined by $h(t)$).
 
Elaborating further the structure of the stochastic perturbation (see  the Appendices), it is possible to
derive the following condition
\begin{equation}
\int_0^\infty G(\tau) \, \left [ k_B \, T \, h(\tau) - C_{RR}(\tau) \right ] \, d \tau=0
\label{eq3_11}
\end{equation}
where $G(t)=L^{-1}[\widehat{G}(s)]$ is the resolvent function defined by eq. (\ref{eq2_6}). 
Eq. (\ref{eq3_11}) is the most general FD2k result  associated 
with the analysis of  the kinetic energy for the initial value GLE. It will
be referred to as the weak formulation of FD2k.
Of course, if
\begin{equation}
C_{RR}(t)= k_B \, T \, h(t)
\label{eq3_12}
\end{equation}
corresponding to the Kubo FD2k for abstract GLE, eq. (\ref{eq3_11})
is identically satisfied, but we cannot  claim eq. (\ref{eq3_12})
directly from eq. (\ref{eq3_11}).

\noindent

{\bf Local realizability - } In order to improve the analysis, a further condition on the GLE should
be posed. Indeed, a solution to this problem is provided by the
assumption of {\em local realizability}, introduced below.
Consider again the memory term  $-h(t) * v(t)$ in eq.  (\ref{eq1_4}) where ``$*$'' stands for convolution, corresponding to
the force  exerted by  the fluid on the particle
at time $t$, and a continuous kernel $h(t)$ (i.e. not impulsive as in the case of the Stokes-Einstein kernel, $h(t)= \eta \, \delta (t)$). This action is local in time, in the meaning that it
 should be
properly viewed as the lumped and compact mathematical
representation of local effects,
occurring instantaneously in time, i.e. at time $t$.  This means that
a vector-valued  $n$-dimensional function ${\bf z}(t)$ of time and a scalar function
$\psi:{\mathbb R}^n \rightarrow {\mathbb R}$ should exist, such that $-h(t)*v(t)=\psi({\bf z}(t))$, where 
${\bf z}(t)$ evolves according to a differential equation
$d {\bf z}(t)/dt = {\bf f}({\bf z}(t),v(t),t)$.

But eq. (\ref{eq1_4}) is linear and it gives rise to a stationary
stochastic process $v(t)$. This implies that the vector field ${\bf f}({\bf z}(t),v(t),t)$
should be  linear in both ${\bf z}(t)$ and $v(t)$ and autonomous (time-independent), and  $\phi({\bf z}(t))$  a linear (and 
homogeneous) functional of ${\bf z}(t)$. 
Consequently,  the memory term entering eq. (\ref{eq1_4}) should be the expressed as
the linear projection of the local functions  ${\bf z}(t)=(z_1(t),\dots,z_n(t))$,
be their system finite or countable,  where ${\bf z}(t)$ satisfies 
an ordinary linear differential equation with constant coefficients driven by stochastic fluctuations, that
 can be expressed as a linear combination
of white noise processes (by eq. (\ref{eq1_2})).

We thus arrive at the definition of local realizability.
The GLE is said to be {\em locally realizable}
if there exist 
a  constant $n \times n$ matrix $\boldsymbol{\Lambda}$ and
a constant $n$-vector ${\bf a}$ such that
eq. (\ref{eq1_4}) for $R(t)=0$ can be expressed as the projection with
respect to ${\bf a}$ of an $n$-dimensional process of internal
degrees of freedom ${\bf z}(t)$,
\begin{equation}
m \, \frac{d v(t)}{d t}= - \sum_{h=1}^n a_i \, z_i(t)= - \left ({\bf a}, {\bf z}(t) \right )
\label{eq3_13}
\end{equation}
evolving according to the local dynamics
\begin{equation}
\frac{d {\bf z}(t)}{d t} = - \boldsymbol{\Lambda} \, {\bf z}(t) + {\bf 1} \, v(t)
\label{eq3_14}
\end{equation}
where ${\bf 1}=(1,\dots,1)$. From the mathematical point of view, local
realizability is one-to-one with the property that $h(t)$ possesses
a finite or countable number of pole singularities.
It follows from this definition that
\begin{equation}
 h(t)= \left ( {\bf a}, e^{-\boldsymbol{\Lambda} \, t} \, {\bf 1} \right )
\label{eq3_15} 
\end{equation}
The GLE is {\em stocastically realizable} if there exists a constant $n \times n$ matrix
$\boldsymbol{\beta}$ such that eq. (\ref{eq1_4})
 can be expressed by eq. (\ref{eq3_13}) with
\begin{equation}
\frac{d {\bf z}(t)}{d t} = - \boldsymbol{\Lambda} \, {\bf z}(t) + {\bf 1} \, v(t)
 + \sqrt{2} \, \boldsymbol{\beta} \, \boldsymbol{\xi}(t)
 \label{eq3_16}
\end{equation}
where $\boldsymbol{\xi}(t)=(\xi_1(t),\dots,\xi_n(t))$ is a vector-valued
white noise process \cite{vankampen}, 
 $\langle \xi_i(t) \, \xi_j(\tau) \rangle = \delta_{ij} \, \delta(t-\tau)$,
and eqs. (\ref{eq3_13}), (\ref{eq3_16}) admit  an equilibrium
invariant measure for which eqs. (\ref{eq1_3}) and (\ref{eq2_5}) hold
(i.e. FD1k is verified).
In practice, one may choose
$\xi_i(t)=d w_i(t)/dt$,
where $w_i(t)$ are independent Wiener processes.

It is easy  to observe that local realizability involves exclusively the
dissipative contribution, while stochastic realizability
corresponds to  the existence
of a Markov embedding of eq. (\ref{eq1_4}),
and specifically the existence of a stochastic process $R(t)$,
 with the property of fulfilling the Langevin
condition eq. (\ref{eq1_2}) and in turn eqs. (\ref{eq1_3}), (\ref{eq2_5}).

Without loss of generality, assume that the internal degrees of freedom $z_i(t)$  are
expressed in the eigenbasis of $\boldsymbol{\Lambda}$ so that
$\boldsymbol{\Lambda}=\mbox{diag}(\lambda_1,\dots,\lambda_n)$.
Assume $\lambda_i \in {\mathbb R}$, with $\lambda_i >0$, that corresponds to the  dissipative
friction factor in viscoelastic fluids \cite{rheol1,rheol2}.
At the moment consider the case where $a_i>0$, $i=1,\dots,n$,  as dictated by the
rheological analysis \cite{rheol1,rheol2}. 
In this case, eq. (\ref{eq3_14}) takes the
 simpler form
    $h(t) = \sum_{i=1}^n a_i \, e^{-\lambda_i \, t}$.

To begin with,  assume that the stochastic fluctuations acting on
the internal models are characterized by the commutative property
$[\boldsymbol{\Lambda},\boldsymbol{\beta}]= \boldsymbol{\Lambda} \boldsymbol{\beta} - \boldsymbol{\beta} \boldsymbol{\Lambda}=0$,
and consequently the matrix
$\boldsymbol{\beta}$ is also diagonal once expressed in  the eigenbasis of $\boldsymbol{\Lambda}$,
$\boldsymbol{\beta}=  \mbox{diag}(c_1,\dots,c_n)$.

Under these hypotheses, eqs (\ref{eq3_12}), (\ref{eq3_16}) admit an equilibrium invariant measure, and
FD1k, i.e. eq. (\ref{eq2_5}), is fulfilled provided that the conditions 
\begin{equation}
\langle z_i \, v \rangle_{\rm eq} =0 \, , \qquad i=1,\dots,n
 \label{eq3_17}
 \end{equation}
 are met (see the Appendices), indicating that the internal variables $z_i(t)$ should be uncorrelated
from the velocity $v(t)$ at equilibrium. Enforcing eq. (\ref{eq1_3}), the  expression for the
  coefficients
$c_i$ follows 
\begin{equation}
 c_i^2 =   m \, \langle v^2 \rangle   
 \, \frac{\lambda_i}{a_i}  \, , \qquad i=1,\dots,n
 \label{eq3_18}
 \end{equation}
consistently with the analysis developed by Goychuk \cite{goychuk}.
Since the 
process  $R(t)$ attains  at equilibrium  the   expression
$R(t)= -  \sqrt{2} \, 
\sum_{i=1}^n a_i \, c_i \,  e^{-\lambda_i \, t} * \xi_i(t)$,
 from eq. (\ref{eq3_16})  it readily follows  that (see the Appendices) 
\begin{equation}
C_{RR}(t)=  
     k_B \, T \, \sum_{i=1}^n a_i e^{-\lambda_i \, t} = k_B \, T \, h(t)
 \label{eq3_19}
 \end{equation}
i.e. that FD2k in its strong form is verified.

\noindent

{\bf Consistency conditions: Thermodynamic equilibria and FD failures - } 
Before considering the case of negative $a_i's$, let us address a preliminary
observation. FD theory applies if a thermodynamic equilibrium state
exists, which is a property that cannot be given for granted for GLE, 
even involving 
``well-behaved'' dissipative kernels.  The  latter property can be defined by the conditions: (i)
 $h(t) \geq 0$; (ii)  $h(t) \in L_1([0,\infty))$, and
thus $0 < \eta_\infty = \int_0^\infty h(t) dt < \infty$;
(iii) $h(t)$ is exponentially decaying with time $t$ (for large $t$).
 In the classical FD theory property (i)+(ii) ensures that the Laplace-Fourier
transform of $h(t)$ exists, and  that  the effective diffusivity $D$ fulfils
the generalized Stokes-Einstein relation $D = \widehat{C}_{vv}(0)= k_B \, T/\eta_\infty$.
But this is not true in general. Let us consider a simple example and derive out of it some 
general conclusions.
Consider a non-dimensional formulation of the GLE which implies that we can set in eq. (\ref{eq1_4})
$m=1$, and $\langle v^2 \rangle_{\rm eq}=1$, and a     
 simple two-mode  kernel
\begin{equation}
h(t)=a_1 \, e^{-\lambda_1 t} + a_2 \, e^{-\lambda_2  t}
\label{eqx_1}
\end{equation}
 with $a_1=1$, $a_2=-\alpha$, $\lambda_1=1/10$, $\lambda_2=1$,
where $\alpha \leq 1$ is taken as  a parameter. 
The graphs of $h(t)$  for several
values of $\alpha$ are depicted in figure \ref{Fig1}.

\begin{figure}
\includegraphics[width=6cm]{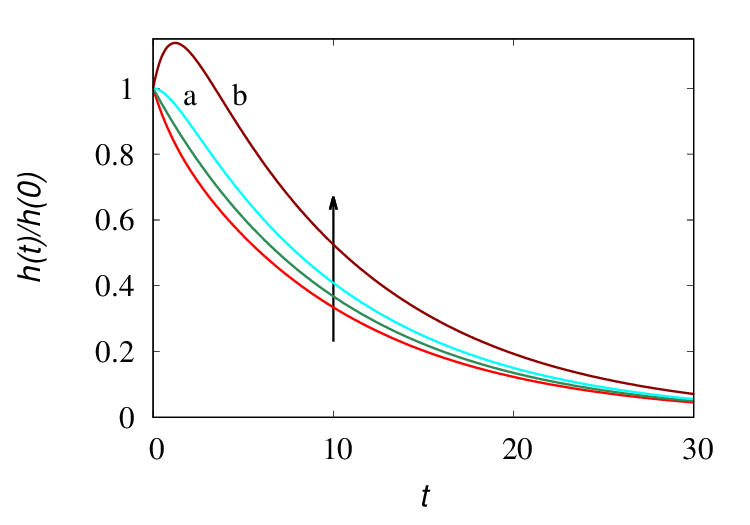}
\caption{$h(t)/h(0)$ vs $t$ for the model considered in
the main text. The arrow indicates increasing values of
$\alpha=-0.1,\,0,\, 0.1,\,0.3$. Line (a) refers to $\alpha=0.1$, line (b) to
$\alpha=0.3$.}
\label{Fig1}
\end{figure}

For these kernels, the conditions (i)-(iii) stated above 
are met and $\eta_\infty =a_1/\lambda_1+a_2/\lambda_2=10-\alpha$.
Setting  ${\bf y}=(v,z_1,z_2)$ the overall dynamics
eqs. (\ref{eq3_13})-(\ref{eq3_16}) can be
compactly expressed as $d {\bf y}/dt= {\bf A} \, {\bf y}+
\sqrt{2} \, \boldsymbol{\beta} \, \boldsymbol{\xi}$, where the coefficient matrix ${\bf A}$
is given by
\begin{equation}
{\bf A} =
\left (
\begin{array}{ccc}
0 & -a_1 & \alpha \\
1 & -\lambda_1 & 0 \\
1 & 0 & -\lambda_2
\end{array}
\right )
\label{eqx_2}
\end{equation}
The eigenvalue spectrum  of ${\bf A}$ controls
the dynamic response of the system. For $\alpha \in (-\infty, \alpha_{\rm ds})={\mathcal D}_{\rm ds}$,
with $\alpha_{\rm ds}=0.21$ all the eigenvalues
possess negative real part: for $\alpha \in (0,\alpha_{\rm ds})$ 
there exists a real eigenvalue and a couple of complex conjugate eigenvalues.
 Above $\alpha_{\rm ds}$, the
real part of the complex-conjugate pair becomes positive,
while the real eigenvalue remains negative.
Therefore,  for $\alpha> \alpha_{\rm ds}$
the stochastic dynamics is unstable and the system does not
attain an equilibrium behavior, as the fluctuations cannot be  restricted to
the stable eigenspace of ${\bf A}$.
We refer to ${\mathcal D}_{\rm ds}=(-\infty,\alpha_{\rm ds})$ as to the region
of {\em dissipative stability} of the GLE. Outside ${\mathcal D}_{\rm ds}$ no thermodynamic
equilibrium exists, because the ``dissipative'' dynamics of the GLE 
 is unstable. 

This result follows  also from the general theory of linear integro-differential
equations  \cite{math1,math2}: the condition for asymptotic stability of eq. (\ref{eq1_4}) is that
$ (  m \, s   +   \widehat{h}(s)) \neq 0$ for $\mbox{Re}[s] \geq 0$, corresponding to
the absence of eigenvalues of ${\bf A}$ with positive real part.
In this regard, the occurrence of instabilities in GLE with ``well-behaved'' kernels goes
beyond the assumption of local realizability of $h(t)$.
For  GLE with ``well-behaved'' kernels that are not dissipatively stable,  the
Fourier-Laplace transform of $h(t)$ is defined outside the half-plane of 
convergence of $\widehat{h}(s)$ (that is $\mbox{Re}[s]>\mu_{\rm max}>0$, with $\mu_{\rm max}>0$ the
largest positive real part of the eigenvalues of ${\bf A}$) and,
as a consequence, it neither describes the response to external 
perturbations (for instance periodic ones)
 nor it permits to derive the expression for the diffusivity 
(that in these situations simply does not exist).
As a further remark, the problem of ergodicity breaking considered in
\cite{bao1,bao2,bao3,plyukhin1,plyukhin2} corresponds to conditions on the kernel lying
at the boundary of the region of dissipative stability \cite{ergbreak}.

We can use further eq. (\ref{eqx_1}) as a model for unveiling the constructive structure of the
FD theory.
From what said above,
 the region $\alpha \in {\mathcal D}_{\rm c}=(-\infty,0] \subset {\mathcal D}_{\rm ds}$,
 and consequently
an equilibrium state exists. Moreover, we have obtained
 constructively that the matrices $\boldsymbol{\Lambda}$ and $\boldsymbol{\beta}$
commute, and that the FD theory applies. Let $\alpha_{\rm c}=0$ be 
the  upper value of $\alpha$ in this region.

For $\alpha \in (\alpha_{\rm c},\alpha_{\rm ds})$ the eigenrepresentation of $\Lambda$ does not diagonalize $\boldsymbol{\beta}$,
and therefore $\boldsymbol{\beta}$ should be  a full matrix. The mathematical details 
are developed in the Appendices, and rely on the 
properties of $\boldsymbol{\beta}$. 
 It is  shown that for $\alpha= (\alpha_{\rm c}, \alpha_{\rm nc})$ with $\alpha_{\rm nc}=0.1$,
a non-diagonal  matrix $\boldsymbol{\beta}$ can always be defined, such  that FD1k and FD2k are
 verified.
 Conversely,
for $\alpha \in (\alpha_{\rm nc}, \alpha_{\rm ds})$   there is no stochastic model  consistent 
with the FD1k eq. (\ref{eq2_5}) although, for any choice of $\boldsymbol{\beta}$, the stochastic dynamics 
eqs. (\ref{eq3_13}),(\ref{eq3_16}) admits a stationary non-equilibrium  invariant measure.
For $\alpha={\mathcal D}_{sr}=(-\infty,\alpha_{nc})$    the GLE is {\em stochastically realizable}  and this means that:
(i)  the GLE is not unstable, and 
(ii) a model for the thermal force  can be realized such
 that the corresponding stationary measures/fluctuations 
 correspond to those in thermal equilibrium  conditions.
It follows from (i) that ${\mathcal D}_{sr} \subseteq {\mathcal D}_{\rm ds}$, (ii)
$\alpha_{\rm c} \leq \alpha_{\rm nc} \leq \alpha_{\rm ds}$, and  consequently
the FD theory applies for kernels belonging to  ${\mathcal D}_{sr}$.
To make a numerical example, figure \ref{Fig2} (upper panel) depicts the velocity autocorrelation functions 
for $\alpha = 0.08 \in {\mathcal D}_{\rm sr}$ (inner dots), and the ``ideal'' Kubo autocorrelation function
$C_{vv}^{(K)}(t)$
solution of eq.  (\ref{eq2_5}) for $\alpha=0.18$ (curve a) outside $ {\mathcal D}_{\rm sr} $ but still inside ${\mathcal
D}_{ds}$ that, from what said above, cannot be realized by any stochastic model
consistent with eqs. (\ref{eq1_3}), (\ref{eq2_5}). The lower panel corresponds to 
$C_{vv}^{(K)}(t)$ for $\alpha=0.3$,
outside ${\mathcal D}_{\rm ds}$, for which $|C_{vv}^{(K)}(t)| \sim e^{\mu_{\rm max} \, t}$ where $\mu_{\rm max} \simeq 0.0214$.

\begin{figure}
\includegraphics[width=6cm]{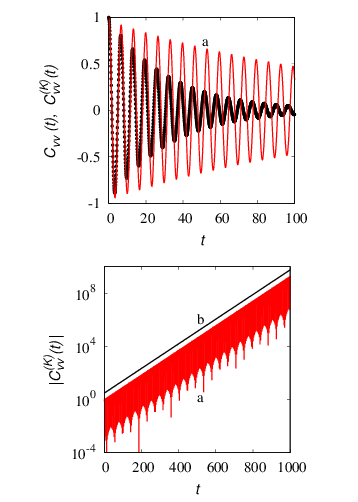}
\caption{Upper panel: $C_{vv}(t)$ (symbols)  for $\alpha=0.08$ and $C_{vv}^{(K)}(t)$ for $\alpha=0.18$ vs $t$.
Lower panel:  $C_{vv}^{(K)}(t)$ (line a) for $\alpha=0.3$ vs $t$. Line (b) corresponds to the
exponential scaling $e^{\mu_{\rm max} \, t}$.} 
 \label{Fig2}
\end{figure}
The properties of ``well-behaved'' kernels in the region  outside ${\mathcal D}_{\rm ds}$ admits relevant implications
in the linear response theory. Specifically, if $h(t)$ possesses real poles with negative real part this
does not  imply that its response to a sinusoidal perturbation  ${\bf F}(\omega)=F_0 e^{\pm \mathrm{i} \omega \, t}$  would be
 bounded and described
by the mobility function $\mu(\omega)=1/(i\omega + h[\omega]$) \cite{kubo2}, where $h[\omega]=\widehat{h}(\mathrm{i} \, \omega)$,
$\mathrm{i}=\sqrt{-1}$. Generically we have an exponentially  diverging   dynamics with
time $t$, where $|v(t)|  \leq   \mu_c^{(e)}(\omega)   \, F_0  \, e^{\mu_{\rm max} \, t}$, if $F(\omega)=F_0 \, \cos(\omega \, t)$, and
 $|v(t)| \leq  \mu_s^{(e)}(\omega)  \, F_0 \,  e^{\mu_{\rm max} \, t}$, if $F(\omega)=F_0 \, \sin(\omega \, t)$,
with obviously $\mu_s^{(e)}(\omega) \neq \mu_c^{(e)}(\omega)$. The graph of these functions is depicted in
figure \ref{Fig3} for $\alpha=0.3$.  In these cases,  i.e. for dissipatively unstable systems, the classical linear response theory
should be revisited. It follows straightforwardly that, outside ${\mathcal D}_{\rm ds}$ no
Stokes-Einstein relations can be defined, albeit $\eta_\infty = \int_0^\infty h(t) \, d t>0$ may exist.

\begin{figure}
\includegraphics[width=6cm]{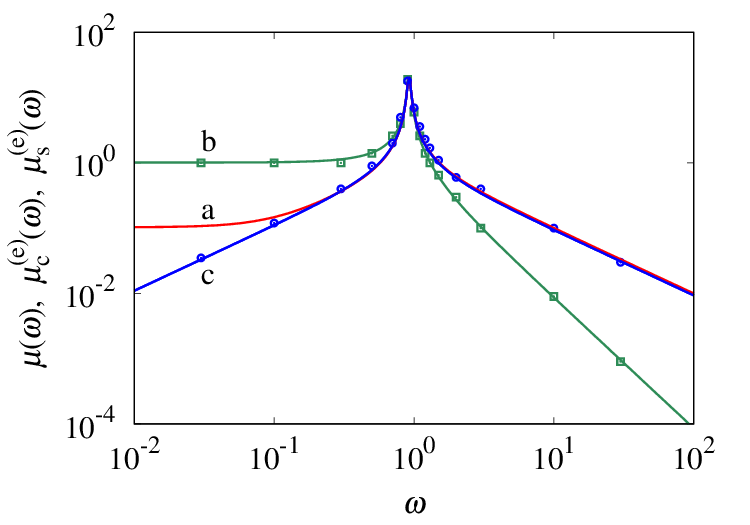}
\caption{ $\mu(\omega)$ line (a), $\mu_c^{(e)}(\omega)$ line (b), and $\mu_s^{(e)}(\omega)$ line (c) predicted analytically (see the Appendices).
Symbols, are the results of numerical  simulations.}
 \label{Fig3}
\end{figure}

{\bf Concluding remarks - }
The  theory above developed provides clear thermodynamic conditions to be set for the memory response
function $h(t)$ of physical equilibrium systems. In order to be thermodynamically consistent, $h(t)$  should be: 
dissipatively
stable, as otherwise it is meaningless  to assume  the occurrence of stationary properties (and a-fortiori the existence
of an equilibrium state), and (ii) stochastically realizable, as otherwise the  steady states
do not correspond to  thermodynamic equilibrium  conditions
driven by stochastic fluctuations. The first property involves
exclusive the  dissipative dynamics, i.e. the kernel $h(t)$. 
The condition of stochastic  realizability provides a constructive and definite limit
to the FD theory. The application of FD theory for systems
outside the region ${\mathcal D}_{sr}$ is not only  mathematically
incorrect, but also  may lead to completely erroneous physical predictions.
Retrospectively,  the ``wormhole'' of  the classical FD theory based on
eq. (\ref{eq1_1}) appears clear: since it assumes the occurrence of thermodynamic equilibrium conditions
without assessing them,
 it fails whenever  the equilibrium conditions cannot occur either by dissipative instability or by the
lack of stochastic realizability. This shortcoming  cannot occur  within the initial-value formulation
of FD theory,
as it is grounded on  a constructive approach to FD2k. This is the principal  methodological difference in the two approaches,
as from the mathematical point of view the conditions assessing  the asymptotic stability of eq. (\ref{eq1_4})
are  fully equivalent  to those  for eq. (\ref{eq1_1}) \cite{jap}.
The analysis of the dissipative  stability of ``well-behaved kernels'', such as those
defined by eq. (\ref{eqx_1}),  indicates  further that the  physical meaning of dissipation is 
much more subtle  either than the existence of a positive and finite value of $\eta_\infty>0$ and $\mbox{Re}[\lambda_i]>0$,
or than the condition $\mbox{trace}({\bf A}) <0$,
as it necessarily involves a description of the internal dynamics in order to assess its thermodynamic
plausibility. 
Moreover, it is possible to show the breakdown of
the constraints of dissipative stability and stochastic realizability ever for
monotonically non-increasing kernels, i.e. such that $d h(t)/d t \leq 0$ for all $t \geq 0$.
This hinges towards a constructive approach to FD theory 
grounded on the explicit representation of $R(t)$ in terms of  stochastics 
processes, as developed above.
A thorough elaboration of the dynamic theory of GLE will be developed in
a longer communication, with particular emphasys to
hydromechancal interactions and Brownian motion
and to
the Debye relaxation mechanism \cite{debye1,debye2}.\\

\vspace{0.1cm}

{\bf Appendix I - Kinetic energy balance -}
Kinetic energy balances represent a convenient way to approach FD2k starting from the initial-value
representation.
To begin with, consider the classical memoryless Langevin model for a Brownian  spherical
particle,  where the thermal force is proportional to the distributional
derivative of a Wiener process $R(t)=\alpha \, d w(t)/dt$,
\begin{equation}
m \, d v(t)= - \eta \, v(t) \, dt + \alpha \, d w(t)
\label{S-eq0_1}
\end{equation}
and $\eta$ is the friction factor. 
Consider the evolution equation
for the kinetic energy. Due to the singular nature of eq. (\ref{S-eq0_1}), the Ito lemma  can be
applied,
\begin{eqnarray}
\frac{m}{2} d v^2(t) & =  & m \, v(t) \, d v(t) + \frac{m}{2} \left ( d v(t)  \right )^2
\nonumber \\
& = & - \eta \, v^2(t) \, d t+ \alpha \, v(t) \, d w(t) + \frac{\alpha^2}{2 \, m} \, dt + o(dt)
\label{S-eq0_2}
\end{eqnarray}
Taking the expected value of both sides of eq. (\ref{S-eq0_2}) with
respect to the probability measure of the fluctuations at equilibrium, enforcing  the Langevin condition,
i.e. $\langle  v(t) \, d w(t) \rangle =0$, and considering that at equilibrium $d \, \langle v^2 \rangle_{\rm eq}=0$,
we obtain to the $dt$-order, $-\eta \, \langle v^2 \rangle + \alpha^2/2 \, m=0$, leading to
$\alpha= \sqrt{2 \, k_B \, T \, \eta}$.

The same  approach can be applied to the GLE.
Due to the convolutional nature of the
dissipation term, it is natural to consider  a convolutional expression for $R(t)$, i.e.,
\begin{equation}
m \, \frac{d v(t)}{d t} = - \int_0^t h(t-\tau) \,  v(\tau) \, d \tau+ \int_0^t \gamma(t-\tau) \, \xi(\tau) \, d \tau
\label{S-eq0_3}
\end{equation}
where $\xi(t)$ is the distributional derivative of a Wiener process, and $\gamma(t)$ a smooth memory kernel. In this case, the
stochastic force is given by $R(t)=\int_0^t \gamma(t-\tau) \, \xi(\tau) \, d \tau$, and 
its autocorrelation function can be expressed as
\begin{equation}
C_{RR}(t)= \langle R(t) \, R(0) \rangle = \int_0^\infty \gamma(t+\tau) \, \gamma(\tau) \, d \tau
\label{S-eq0_4}
\end{equation}

Similarly to
the impulsive case, one can to derive FD2k
by considering the evolution  of the kinetic energy. From eq. (\ref{S-eq0_3}) we obtain
\begin{eqnarray}
\frac{m}{2} \, \frac{d \langle v^2(t) \rangle}{d t} & = & - \int_0^t h(t-\tau) \, \langle v(\tau) \, v(t) \rangle
\, d \tau + \int_0^t \gamma(t-\tau) \, \langle   \xi(\tau) \, v(t) \rangle \, d \tau
\nonumber \\
& = & - \int_0^t h(t-\tau) \, C_{vv}(t-\tau) \, d \tau + \int_0^t \gamma(t-\tau) \,
C_{v \xi}(t-\tau) \, d \tau
\label{S-eq0_5a} \\
& = & - \int_0^t h(\tau) \, C_{vv}(\tau) \, d \tau + \int_0^t \gamma(\tau) \, C_{v \xi}(\tau) \, d \tau
\nonumber
\end{eqnarray}
that has been derived by enforcing the property that $v(t)$ is a stationary stochastic process.
At equilibrium, i.e.  for $t \rightarrow \infty$, the l.h.s. of eq. (\ref{S-eq0_5a}) is vanishing, and this implies
\begin{equation}
\int_0^\infty h(\tau) \, C_{vv}(\tau) \, d \tau= \int_0^\infty \gamma(\tau) \, C_{v \xi}(\tau) \, d \tau
\label{S-eq0_5}
\end{equation}
where $C_{vv}(t)=\langle v(t) \, v(0) \rangle_{\rm eq}$ and $C_{v \xi}(t)=\langle v(t) \, \xi(0) \rangle_{\rm eq}$ 
are the correlation functions at equilibrium
conditions. The evolution equation for   $C_{v\xi}(t)$ follows from eq. (\ref{S-eq0_3}), multiplying it by $\xi(0)$ and
taking the expect value 
\begin{equation}
m \, \frac{d C_{v \xi}(t)}{d t}  = -  \int_0^t h(t-\tau) \, C_{v \xi}(\tau) \, d \tau  + \gamma(t)
\label{S-eq0_6}
\end{equation}
It is  equipped with the initial  condition $C_{v \xi}(0)=0$. Thus,
\begin{eqnarray}
C_{vv}(t)& = &  \langle v^2 \rangle \, G(t) \nonumber \\
C_{v \xi}(t) & = & \int_0^t G(t-\tau) \, \gamma(\tau) \, d \tau=  \frac{1}{m} \, G(t) * \gamma(t)
\label{S-eq0_7}
\end{eqnarray}
where $G(t)$ is the Green function expressed by the inverse Laplace transform
\begin{equation}
G(t)= L^{-1} \left [ \frac{1}{s  + \widehat{h}(s)/m} \right ]
\label{S-eq0_8}
\end{equation}
 Substituting eq. (\ref{S-eq0_7}) into
eq. (\ref{S-eq0_5}) we have
\begin{equation}
m \, \langle v^2 \rangle \int_0^\infty h(\tau) \,  G(\tau) \, d \tau =
\underbrace{\int_0^\infty \gamma(\tau) \, \left [\gamma * G(\tau) \right ] \, d \tau}_{I}
\label{S-eq0_9}
\end{equation}
The integral $I$ at the r.h.s of eq. (\ref{S-eq0_9}) can be expressed as
\begin{eqnarray}
I &= & \int_0^\infty \int_0^\tau \gamma(\tau-\theta) \, G(\theta) \, d \theta  =
\int_0^\infty G(\theta) \, d \theta \int_\theta^\infty \gamma(\tau) \, \gamma(\tau-\theta) \, d \tau
\nonumber \\
& = & \int_0^\infty G(\theta) \, \int_0^\infty \gamma(\tau+\theta) \, \gamma(\tau) \, d \tau =
\int_0^\infty G(\theta) \, C_{RR}(\theta) \, d \theta
\label{S-eq0_10}
\end{eqnarray}
and therefore  eq. (\ref{S-eq0_9})  takes the final form
\begin{equation}
\int_0^\infty G(\tau) \, \left [ k_B \, T \, h(\tau) - C_{RR}(\tau) \right ] \, d \tau=0
\label{S-eq0_11}
\end{equation}
corresponding to the weak formulation of the FD2k reported in the main text.\\

{\bf Appendix II - Stochastic realizations: the commutative case - }
This case corresponds to the mutual diagonalization of the dissipative and 
fluctuational contributions, (i.e. of the
matrices $\boldsymbol{\Lambda}$ and $\boldsymbol{\beta}$).
 Here we outline the basic methodology used, while for
the explicit calculations we refer to the next section, dealing with the
more general condition $[\boldsymbol{\Lambda},\boldsymbol{\beta}] \neq 0$.
Consider the expression for the kernel $h(t)$

\begin{equation}
h(t)=\sum_{i=1}^n a_i \, e^{-\lambda_i \, t}
\label{S-eq1_1}
\end{equation}
with $\lambda_i >0$. To this kernel, assuming $[\boldsymbol{\Lambda},\boldsymbol{\beta}]=0$,
corresponds the local representation of the GLE
\begin{eqnarray}
\frac{d v(t)}{d t} &= & - \sum_{i=1}^n a_i \, z_i(t) \nonumber \\
\frac{d z_i(t)}{d t} & = &- \lambda_i \, z_i(t) + v(t) + \sqrt{2} \, c_i \, \xi_i(t) \, ,  \quad i=1,\dots,n
\label{S-eq1_2}
\end{eqnarray}
where $\xi_i(t)$ can be taken as distributional derivatives of independent
Wiener processes $w_i(t)$,
$\xi_i(t)=d w_i(t)/dt$, $i=1,\dots,n$. In this way, the associated Fokker-Planck equation for
the probability density $p(v,{\bf z},t)$, ${\bf z}=(z_1,\dots,z_n)$ is
parabolic. The moments
\begin{eqnarray}
\langle v^2(t) \rangle & = & \int v^2 \, p(v,{\bf z},t) \, d v d {\bf z} \nonumber \\
\langle z_i(t) \, v(t)  \rangle & = & \int   z_i \, v \, p(v,{\bf z},t) \, d v d {\bf z} = y_i(t) \, ,  \quad i=1,\dots,n 
\label{S-eq1_2a} \\
\langle z_i(t) \, z_j(t)  \rangle & = & \int z_i \, z_j \, p(v,{\bf z},t) \, d v d {\bf z} = m_{i,j}(t) \, , \quad i,j=1,\dots,n 
\nonumber
\end{eqnarray}
can be defined, where
 $d {\bf z}= d z_1 \cdots d z_n$, and their evolution equations
derived, equipped with  arbitrary
initial conditions associated with the initial preparation of the process, defined
e.g.  by any initial density $p_0(v,{\bf z})$, $p(v,{\bf z},t=0)=p_0(v,{\bf z})$.
At equilibrium, if an equilibrum exists, the steady-state moment values  are given by
the equilibrium quantities $\langle v^2 \rangle_{\rm eq}$,  $\langle z_i \, v \rangle_{\rm eq}$,
 $\langle z_i \, z_j \rangle_{\rm eq}$, respectively.

Similarly, by enforcing the Langevin condition, it follows for the correlation
functions $C_{vv}(t)$ and $C_{z_i v}(t)$ at equilibrium,
\begin{eqnarray} 
C_{vv}(t) & = & \langle v(t) \, v(0) \rangle_{\rm eq}= \lim_{\tau\rightarrow \infty}
\langle (v(t+\tau) \, v(\tau) \rangle \nonumber \\
C_{z_i v}(t) & = & \langle z_i(t) \, v(0) \rangle_{\rm eq}= \lim_{\tau\rightarrow \infty}
\langle (z_i(t+\tau) \, v(\tau) \rangle \, , \qquad i=1,\dots,n
\label{S-eq1_3}
\end{eqnarray}
the system of evolution equations
\begin{eqnarray}
\frac{d C_{vv}(t)}{d t} & = & - \sum_{i=1}^n a_i \, C_{z_i v}(t) \nonumber \\
\frac{d C_{z_i v}(t)}{d t} & = & - \lambda_i \, C_{z_i v}(t) + C_{vv}(t)
\label{S-eq1_4}
\end{eqnarray}
equipped with the initial conditions,
\begin{equation}
C_{vv}(0)= \langle v^2 \rangle_{\rm eq} \, ,  \qquad C_{z_i v}(0)= \langle z_i \, v \rangle_{\rm eq}
\, , \quad i=1,\dots, n
\label{S-eq1_5}
\end{equation}
From the second  system of equations (\ref{S-eq1_4}) we have
\begin{equation}
C_{z_i v}(t)= \langle z_i \, v \rangle_{\rm eq} \, e^{-\lambda_i \, t} + e^{-\lambda_i \, t} * v(t) \, ,
\qquad i=1,\dots,n
\label{S-eq1_6}
\end{equation}
where we use the symbol ``*'' to indicate the convolution integral between any two  functions 
$f(t) * g(t)= \int_0^t f(t-\tau) \, g(\tau) \, d \tau$.
Substituting eqs. (\ref{S-eq1_6}) into the first eq. (\ref{S-eq1_4}) we obtain
\begin{equation}
\frac{d C_{vv}(t)}{d t} = - h(t) * C_{vv} (t) - \sum_{i=1}^n e^{-\lambda_i \, t}
\langle z_i \, v \rangle_{\rm eq}
\label{S-eq1_7}
\end{equation}
In order to recover the fluctuation-dissipation relation of the first kind (FD1k), that
is a consequence of the Langevin condition, we should have identically
\begin{equation}
\langle z_i \, v \rangle_{\rm eq}=0 \, , \qquad i=1,\dots,n
\label{S-eq1_8}
\end{equation}
In the initial-value representation of the FD theory, the system of equations (\ref{S-eq1_8})
represents the fundamental constraint imposed on the internal degrees of freedom, out of
which the equilibrium FD theory can be developed. Specifically,
from eq. (\ref{S-eq1_8}), the expression for the intensity coefficients $c_i$  directly follows  after elementary algebraic manipulations of the steady-state
moment equations.\\

{\bf Appendix III - Stochastic realizations: the noncommutative case - }
In the main text we have considered the following kernel
\begin{equation}
h(t)=a_1 \, e^{-\lambda_1 \, t} + a_2 \, e^{-\lambda_2 \, t}
\label{S-eq2_1}
\end{equation}
in nondimensional form (i.e. $m=1$ for the mass, and $\langle v^2 \rangle_{\rm eq}=1$ for the velocity),
with $a_1=1$, $a_2=-\alpha$, $\lambda_1=1/10$, $\lambda_2=1$. For positive values of $\alpha$,
$\boldsymbol{\Lambda}$ and $\boldsymbol{\beta}$ cannot commute to produce an equilibrium behavior.
Therefore,\ $\boldsymbol{\beta}$ should  be a full matrix,
 and the local representation  follows
\begin{eqnarray}
\frac{d v(t)}{d t} & = & - a_1 \, z_1(t) - a_2 \, z_2(t) \nonumber \\
\frac{d z_1(t)}{d t} & = & -\lambda_1 \, z_1(t) + v(t) + \sqrt{2} \, \left ( \beta_{1,1} \, \xi_1(t) +
\beta_{1,2} \, \xi_2(t)   \right )
\label{S-eq2_2} \\
\frac{d z_2(t)}{d t} & = & -\lambda_2 \, z_2(t) + v(t) + \sqrt{2} \, \left ( \beta_{2,1} \, \xi_1(t) +
\beta_{2,2} \, \xi_2(t)   \right )
\nonumber 
\end{eqnarray}
Observe that the stochastic forcing cannot act directly on the velocity variable,
as otherwise no equilibrium conditions could be achieved.

The Fokker-Planck equation for the density $p(v,z_1,z_2,t)$ associated with eq. (\ref{S-eq2_2}) is 
\begin{equation}
\frac{\partial p}{\partial t} = (a_1 \, z_1+ a_2 \, z_2 ) \frac{\partial p}{\partial v} + \sum_{i=1}^2 \lambda_i \frac{\partial
( z_i \, p )}{\partial z_i} - v \, \sum_{i=1}^2 \frac{\partial  p}{\partial z_i} + \sum_{i=1}^2 \sum_{j=1}^2 S_{i,j}
\, \frac{\partial^2 p}{\partial z_i \partial z_j}
\label{S-eq2_3}
\end{equation}
where $\boldsymbol{\beta}=(\beta_{i,j})_{i,j=1,2}$, and the matrix ${\bf S}=(S_{i,j})_{i,j=1,2}$ is defined
by
\begin{equation}
{\bf S}= \boldsymbol{\beta} \, \boldsymbol{\beta}^T
\label{S-eq2_4}
\end{equation}
where  the superscript ``$T$'' indicate transpose. The  matrix ${\bf S}$ should
be  positive definite.
Componentwise,
\begin{equation}
S_{1,1} = \sum_{i=1}^2 (\beta_{1,i})^2 \, , \qquad S_{2,2} = \sum_{i=1}^2 (\beta_{2,i})^2  \, ,
\quad S_{1,2} = S_{2,1}= \sum_{i=1}^2 \beta_{1,i}  \, \beta_{2,i}
\end{equation}
In this case, the moment equations read
\begin{eqnarray}
\frac{d \langle v^2 \rangle}{d t} &=& -a_1 \, y_1 -a_2 y_2 
\label{S-eq2_5} \\
\frac{d y_i}{d t}& =& - \sum_{j=1}^2 a_j m_{i,j} - \lambda_i \, y_i + \langle v^2 \rangle \, ,
\quad i=1,2 
\label{S-eq2_6} \\
\frac{d m_{i,j}}{d t} &= & - (\lambda_i+ \lambda_j) \, m_{i,j}+ y_i + y_j + S_{i,j} + S_{j,i} \, ,
\quad i,j=1,2
\label{S-eq2_7}
\end{eqnarray}
where $y_i(t)= \langle z_i(t) \, v(t) \rangle$, $m_{i,j}(t) = \langle z_i(t) \, z_j(t) \rangle$, $i,j=1,2$.
Imposing the  equilibrium and  the extended Langevin conditions eq. (\ref{S-eq1_8}), i.e.,
 $\langle v^2 \rangle= \langle v^2 \rangle_{\rm eq}=1$, $y_1=y_2=0$,
eq. (\ref{S-eq2_5}) is identically satisfies, while from eq. (\ref{S-eq2_7}) we get
\begin{equation}
m_{1,1}= \frac{S_{1,1}}{\lambda_1} \, , \quad m_{1,2}=m_{2,1}= \frac{ 2 \, S_{1,2}}{\lambda_1+\lambda_2}
\, , \quad m_{2,2}= \frac{S_{2,2}}{\lambda_2}
\label{S-eq2_8}
\end{equation}
Inserting  the latter expressions for $m_{i,j}$ within eqs. (\ref{S-eq2_6}), a  linear system in the ${\bf S}$-matrix
entries is obtained
\begin{eqnarray}
a_1 \, \frac{S_{1,1}}{\lambda_1}+  2 \, a_2  \, \frac{S_{1,2}}{\lambda_1+\lambda_2} &=  & 1
\nonumber \\
 2 \, a_1   \, \frac{S_{1,2}}{\lambda_1+\lambda_2} + a_2 \, \frac{S_{2,2}}{\lambda_2} & = & 1
\label{S-eq2_9}
\end{eqnarray}
If $a_2>0$  (negative $\alpha$), we recover the commutative case, simply setting $S_{1,2}=0$.
For $a_2<0$,   from eq. (\ref{S-eq2_9}) we can explicit the diagonal terms of the ${\bf S}$-matrix
\begin{equation}
S_{1,1} =  \frac{\lambda_1}{a_1} \, \left ( 1 -  \frac{2 \, a_2}{\lambda_1+\lambda_2} \, S_{1,2} \right ) \, , \qquad
S_{2,2} =  -\frac{\lambda_2}{a_2} \left ( \frac{2 \, a_1}{\lambda_1+\lambda_2} \, S_{1,2} - 1 \right )
\label{S-eq2_10}
\end{equation}
Since $-\lambda_2/a_2 >0$, the condition $S_{1,2}> (\lambda_1+\lambda_2)/2 \,a_1$ 
implies that $S_{1,1}>0$, $S_{2,2}>0$. In order to ensure the positive definite nature  of ${\bf S}$
it remains to   fulfil the condition on its determinant, namely $S_{1,1} \, S_{2,2} - \left ( S_{1,2} \right )^2 \geq 0$,
that mathematically can be regarded   as a consequence of the Cauchy-Schwarz inequality.
This leads to the following condition expressed in terms of $S_{1,2}$
\begin{equation}
\frac{\lambda_1}{a_1} \, \left ( 1 -  \frac{2 \, a_2}{\lambda_1+\lambda_2} \, S_{1,2} \right ) 
\, \left (  -\frac{\lambda_2}{a_2}  \right ) \left ( \frac{2 \, a_1}{\lambda_1+\lambda_2} \, S_{1,2} - 1 \right )
- S_{1,2}^2 \geq 0
\label{S-eq2_11}
\end{equation}
Setting $\xi=S_{1,2}$,  and thus $\xi > \xi_{\rm min}=(\lambda_1+\lambda_2)/2 a_1$, eq. (\ref{S-eq2_11}) can be compactly rewritten as a quadratic inequality in $\xi$
\begin{equation}
\phi(\xi)= -1 + (\gamma_2-\gamma_1) \, \xi - \left ( \gamma_3- \gamma_1 \, \gamma_2 \right ) \, \xi^2 \geq 0
\label{S-eq2_12}
\end{equation}
where
\begin{equation}
\gamma_1 = - \frac{2 \, a_2}{\lambda_1+\lambda_2} \, , \quad \gamma_2= \frac{2 \, a_1}{\lambda_1+\lambda_2} \, ,
\quad \gamma_3= - \frac{a_1 \, a_2}{\lambda_1 \, \lambda_2}
\label{S-eq2_13}
\end{equation}
Since
\begin{equation}
\gamma_3-\gamma_1 \, \gamma_2 = - \frac{a_1 \, a_2 \, (\lambda_1-\lambda_2)^2}{\lambda_1 \, \lambda_2 (\lambda_1+\lambda_2)^2}
> 0
\label{S-eq2_14}
\end{equation}
the  coefficient of the quadratic term in $\xi$ of $\phi(\xi)$ is negative. Consequently, 
in order to have stochastic  realizability, the local maximum of $\phi(\xi)$  at $\xi^*$ should
be positive, $\phi(\xi^*)>0$. 
From the expression for $\phi(\xi)$ eq. (\ref{S-eq2_12}), we have
\begin{equation}
\xi^* = \frac{\gamma_2 - \gamma_1}{2 \, (\gamma_3 -\gamma_1 \, \gamma_2 )} \, , \qquad
\phi(\xi^*) = \frac{(\gamma_1+\gamma_2)^2 - 4 \, \gamma_3}{4 \, (\gamma_3  - \gamma_1 \, \gamma_2)}
\label{S-eq2_15}
\end{equation}
Therefore, if $(\gamma_1+\gamma_2)^2 - 4 \, \gamma_3>0$, the GLE is stochastically realizable, and one
can consider any value of $\xi$ falling in the interval $(\xi_-,\xi_+)$, where
\begin{equation}
\xi_{\pm} = \frac{(\gamma_2-\gamma_1)}{2 \,(\gamma_3- \gamma_1 \, \gamma_2)} \, \left [
1 \pm \sqrt{1- \frac{4 (\gamma_3- \gamma_1\,\gamma_2)}{(\gamma_2-\gamma_1)} } \right ]
\label{S-eq2_16}
\end{equation}
provided that these values are greater than $\xi_{\rm min}=(\lambda_1+\lambda_2)/2 \, a_1$ (as is in the
present case).
Figure \ref{Fig1} depicts the behavior of $\phi(\xi^*)$ for $a_1=1$, $\lambda_1=0.1$ $a_2=-\alpha$, $\lambda_2=1$, indicating that $\phi(\xi^*)=0$ for
$\alpha= \alpha_{nc}=0.1$.
\begin{figure}
\includegraphics[width=10cm]{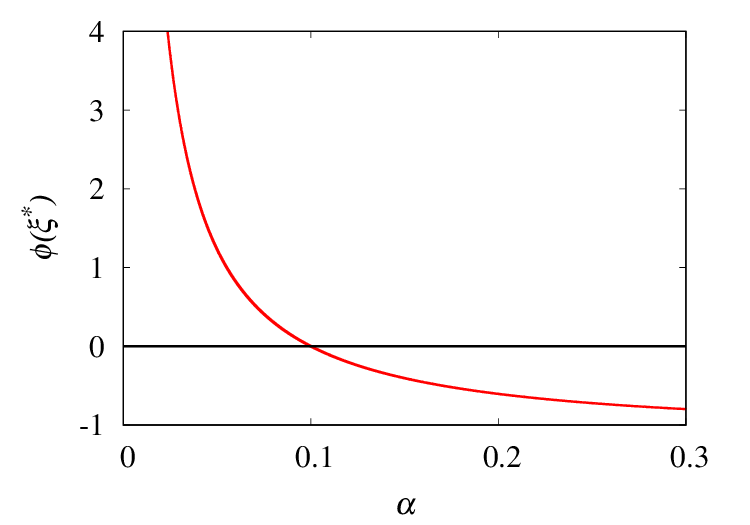}
\caption{ $\phi(\xi^*)$ vs $\alpha$ for $\alpha_1=1$, $\lambda_1=0.1$, $\alpha_2=-\alpha$, $\lambda_2=1$.}
\label{Fig1}
\end{figure}
Figure \ref{Fig2} depicts the behavior of $\xi_+$ and $\xi_-$ vs $\alpha$ in the same conditions.
\begin{figure}
\includegraphics[width=10cm]{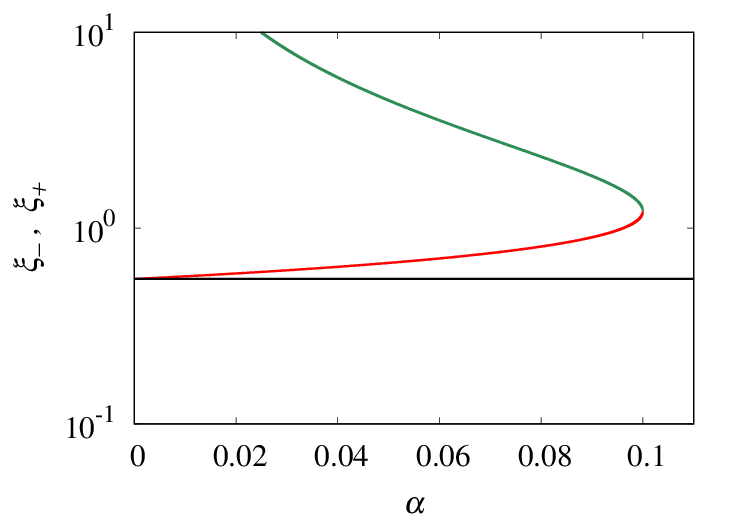}
\caption{$\xi_+$ (upper curve) and $\xi_-$ (lower curve) vs $\alpha$, for $\alpha_1=1$, $\lambda_1=0.1$, $\alpha_2=-\alpha$, $\lambda_2=1$. The horizontal line
corresponds to the value of $\xi_{\rm min}$.}
\label{Fig2}
\end{figure}
Given $\alpha<\alpha_{\rm nc }$,  any value of $\xi \in (\xi_-,\xi_+)$ produces a stochastic realization of the system
at equilibrium, satisfying either  FD1k or  FD2k in its strong form.

For $\alpha < \alpha_{nc}$, any value of $\xi$ in the interval $(\xi_-,\xi_+)$  produces an equivalent model of  the equilibrium
behavior.  To this end, we  have  to show that all these realizations for $\xi \in (\xi_-,\xi_+)$ give rise to one and the same
autocorrelation function  of the resulting  stochastic force at equilibrium (i.e. in the long-term).
To avoid misunderstandings, it is useful to remain that  we  are here considering a nondimensional formulation 
with $m=1$, and $\langle v^2 \rangle_{\rm eq}=1$,
and thus the FD2k result in the strong form corresponds to $C_{RR}(t)= h(t)$.
To verify this property consider the expression for the stochastic force $R(t)$. From eq. (\ref{S-eq2_2})
we have
\begin{equation}
R(t)= \sum_{i=1}^2 a_i \, z_{i,0} \, e^{\-\lambda_i \, t} + \sqrt{2} \sum_{i=1}^2   a_i \sum_{h=1}^2  \, \beta_{i,h} \, e^{-\lambda_i t} *
\xi_h(t) = 2 \sum_{i=1}^2  a_i \sum_{h=1}^2 \beta_{i,h} \, r_{i,h}(t)
\label{S-eq2_17}
\end{equation}
where we have set
\begin{equation}
r_{i,h}(t)=  e^{-\lambda_i t} *
 \xi_h(t) \, , \qquad i,h=1,2
\label{S-eq2_18}
\end{equation}
In the long-term (equilibrium), the first term at the r.h.s. of eq. (\ref{S-eq2_17}) decays exponentially to zero, so
that
\begin{equation}
C_{RR}(t)= \langle R(t) \, R(0) \rangle_{\rm eq}= 2  \sum_{i=1}^2 a_i \sum_{h=1}^2 \beta_{i,h} \sum_{j=1}^2 a_j \sum_{k=1}^2
\beta_{j,k} \, \langle r_{i,h}(t) \, r_{j,k}(0) \rangle_{\rm eq}
\label{S-eq2_19}
\end{equation}
where
\begin{equation}
\langle r_{i,h}(t) \, r_{j,k}(0) \rangle_{\rm eq} =
\lim_{\tau \rightarrow \infty} \langle r_{i,h}(t+\tau) \, r_{j,k}(\tau) \rangle
\label{S-eq2_20}
\end{equation}
The latter expression can be easily calculated by quadraturae, 
enforcing the property $\xi_h(t) \, \xi_k(\tau) = \delta_{h,k} \, \delta(t-\tau)$, and this yields
\begin{equation}
\langle r_{i,h}(t) \, r_{j,k}(0) \rangle_{\rm eq}  = \frac{e^{\lambda_i \, t}}{\lambda_i + \lambda_j} \, \delta_{h,k}
\label{S-eq2_21}
\end{equation}
Inserting the latter expression into eq. (\ref{S-eq2_19}),  we finally arrive to
\begin{eqnarray}
C_{RR}(t) & = & 2  \sum_{i=1}^2 a_i \sum_{h=1}^2 \beta_{i,h} \sum_{j=1}^2 a_j \sum_{k=1}^2
    \beta_{j,k} \,  \frac{e^{\lambda_i \, t}}{\lambda_i + \lambda_j} \, \delta_{h,k}
= 2 \sum_{i=1}^2 a_i \sum_{j=1}^2 \frac{a_j }{\lambda_i + \lambda_j} \, e^{-\lambda_i \, t} \sum_{h=1}^2 \beta_{i,h} \,
\beta_{j,h}  \nonumber \\
& = & \sum_{i=1}^2 a_i \, e^{-\lambda_i \, t}  \, \left [ \sum_{j=1}^1 2 \, a_j \frac{S_{i,j}}{\lambda_i + \lambda_i} \right ]
\label{S-eq2_22}
\end{eqnarray}
The expression within square parentheses at the r.h.s. of eq. (\ref{S-eq2_22}) is identically
equal to one for $i=1,2$ because of eqs. (\ref{S-eq2_9}), and thus eq. (\ref{S-eq2_22}) simply  becomes
\begin{equation}
C_{RR}(t) = \sum_{i=1}^2 a_i \, e^{-\lambda_i \, t}  = h(t)
\label{S-eq2_23}
\end{equation}
which proves that for any plausible choice of $\xi$, all the stochastic realizations are equivalent and satisfy FD2k.\\

{\bf Appendix IV - Linear response theory: exponential divergence outside ${\mathcal D}_{\rm ds}$ - }
Consider the case of ``well-behaved kernels'' $h(t)$ possessing  real and negative poles.
In this case, the classical theory predicts   a bounded response  in the presence of   a  periodically forced  term \cite{kubo2}
\begin{equation}
\frac{d v(t)}{d t}= -\int_0^t h(t-\tau) \, v(\tau) \, d \tau + F(t;\omega)
\label{S-eq3_1}
\end{equation}
where $F(t;\omega)$ is a sinusoidal perturbation of frequency $\omega$, say $ F(t;\omega)= F_0 \, e^{\mathrm{i} \, \omega \, t}$,
i.e., $v(t;\omega)= \mu(\omega) \, F_0 \, e^{\mathrm{i} \, \omega \, t}$, controlled
by the mobility function $\mu(\omega)$ \cite{kubo2},
\begin{equation}
\mu(\omega)= \frac{1}{\mathrm{i} \, \omega+ h[\omega]}
\label{S-eq3_2}
\end{equation}
where $h[\omega]=\widehat{h}(\mathrm{i} \, \omega)$ is the Laplace-Fourier transform of $h(t)$.
This is not true if the GLE is not dissipatively stable, for which the dynamics  diverges exponentially in time
in the presence of  generic sinusoidal perturbations,
 with an exponent controlled by
the  real part of  dominant eigenvalue, i.e. of the eigenvalue possessing the largest real part.

\begin{figure}
\includegraphics[width=10cm]{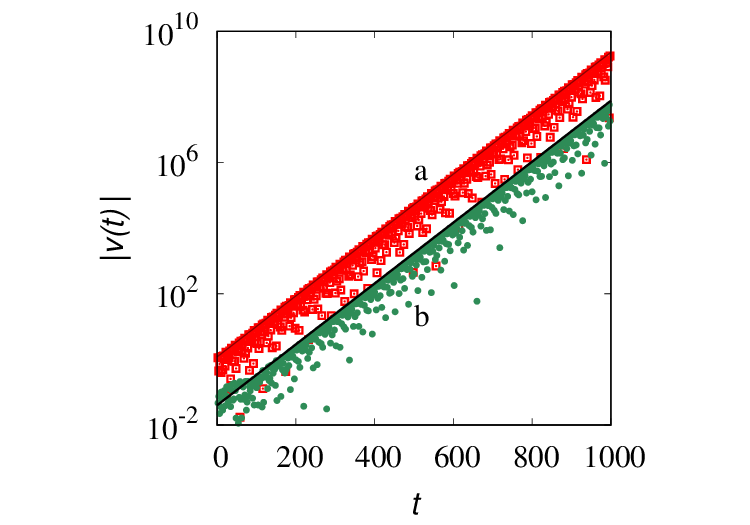}
\caption{$| v(t)|$ vs $t$ for sinusoidal forcings: (a): $F(t,;\omega)=\cos(\omega \, t)$; (b): $F(t;\omega)=\sin(\omega \, t)$ at $\omega=0.03$.}
\label{Fig3}
\end{figure}

To show this, consider the kernel $h(t)$ given by eq. (\ref{S-eq2_1}) at $\alpha=0.3$. In this
case, the eigenvalues of the matrix ${\bf A}$ associated with $h(t)$ are:  $\mu_{1,2}=\mu_u \pm \mathrm{i} \, \omega_u$,
with $\mu_u \simeq 2.137 \, \times 10^{-2}$,  $\omega_u \simeq 0.921$, and $\mu_3=\mu_s \simeq -1.14274$,
corresponding to a couple of unstable complex conjugate  eigenvalues and a stable  real one.
The response to any perturbation can be easily estimated in the Laplace domain  (assuming $v(t=0)=0$),
\begin{eqnarray}
\widehat{v}(s)& = & \frac{1}{s+\widehat{h}(s)} \, \widehat{F}(s;\omega) \nonumber \\
& = & \frac{(s+\lambda_1) \, (s+\lambda_2)}{s \, (s+\lambda_1) \, (s+\lambda_2) +a_1 \, (s+\lambda_2)+ a_2 \,
(s+\lambda_1)} \, \widehat{F}(s;\omega) 
\label{S-eq3_3} \\
& = & \left [  \frac{A + B \, s}{(s-\mu_u)^2 + \omega_u^2} + \frac{C}{s-\mu_s} \right ] \,  \widehat{F}(s;\omega)
\nonumber
\end{eqnarray}
where $A,\,B\,, C$ can be easily calculated from the expansion in partial fractions.
For large $t$,  solely the first term at the r.h.s. of eq. (\ref{S-eq3_3}) is relevant and it
corresponds to a  $v(t)$  exponentially growing in time with the exponent $\mu_{\rm max}=\mu_u$
with a prefactor that depends on the frequency  $\omega$. Specifically
\begin{equation}
v(t) = \mu_c^{(e)}(\omega) \, F_0   \, e^{\mu_u \, t}   \, \cos(\omega_u \, t+\varphi_c)
\label{S-eq3_4}
\end{equation}
if $F(t;\omega)=F_0 \, \cos(\omega \, t)$, and
\begin{equation} 
v(t) = \mu_s^{(e)}(\omega) \, F_0   \, e^{\mu_u \, t}   \, \cos(\omega_u \, t+\varphi_s)
\label{S-eq3_5}
\end{equation}
if $F(t;\omega)=F_0 \, \sin(\omega \, t)$, where 
$\mu_c^{(e)}(\omega)$ and $\mu_s^{(e)}(\omega)$, referred to as the ``exponential mobilities'' can be obtained
analytically from eq. (\ref{S-eq3_3}) in the two cases considered, and  $\varphi_c$ and $\varphi_s$ are the phase-shifts. 
Figure \ref{Fig3} shows some examples of this exponential behavior, while the
behavior of the  exponential mobilities $\mu_c^{(e)}(\omega)$ and $\mu_s^{(c)}(\omega)$ is reported in the main 
text. For dissipatively unstable systems, apart from the exponential divergence with time,
 the frequency of  the oscillations is locked to $\omega_u$, independently of $\omega$,
marking the difference with respect to the dissipatively stable case.

\end{document}